\documentstyle[aps,prd,latexsym,epsfig]{revtex}
\tighten

\newcommand{\HH}{{\cal H}}

\newcommand{\al}{\alpha}

\newcommand{\de}{\delta}

\newcommand{\ga}{\gamma}
\newcommand{\Ga}{\Gamma}
\newcommand{\ka}{\kappa}

\newcommand{\La}{\Lambda}
\newcommand{\la}{\lambda}
\newcommand{\Om}{\Omega}

\newcommand{\ra}{\rightarrow}

\newcommand{\be}{\begin{equation}}
\newcommand{\ee}{\end{equation}}

\newcommand{\lsim}{\stackrel{<}{\sim}}
\newcommand{\bea}{\begin{eqnarray}}
\newcommand{\eea}{\end{eqnarray}}
\newcommand{\bean}{\begin{eqnarray*}}
\newcommand{\eean}{\end{eqnarray*}}
\newcommand{\dd}{\partial}

\newcommand{\etal}{{\em et al.}}

\def\spose#1{\hbox to 0pt{#1\hss}}
\def\ltapprox{\mathrel{\spose{\lower 3pt\hbox{$\mathchar"218$}}
\raise 2.0pt\hbox{$\mathchar"13C$}}}
\def\gtapprox{\mathrel{\spose{\lower 3pt\hbox{$\mathchar"218$}}
\raise 2.0pt\hbox{$\mathchar"13E$}}}

\newcommand{\EI}{\small{\rm{E}}}

\begin{document}

%\draft
%\preprint{\
%\begin{tabular}{rr}
%CfPA/96-th-15 &  \\
%&
%\end{tabular}
%}
\twocolumn[\hsize\textwidth\columnwidth\hsize\csname@twocolumnfalse\endcsname
\title{Adiabatic perturbations in pre-big bang models:
\\ matching conditions and scale invariance}
\author{Ruth Durrer and  Filippo Vernizzi}
\address{
D\'epartement de Physique Th\'eorique, Universit\'e de Gen\`eve,
24 quai E.\ Ansermet, CH-1211 Gen\`eve 4, Switzerland
}

\maketitle

\begin{abstract}
At low energy, the four-dimensional effective action
 of the ekpyrotic model of the universe is equivalent to a
slightly modified version of the pre-big bang model. We discuss
cosmological perturbations in these models. In  particular we
address the issue of matching the perturbations from a collapsing
to an expanding phase. We show that, under certain physically
motivated and quite generic assumptions on the high energy
 corrections, one obtains $n=0$ for the
spectrum of scalar perturbations in the original pre-big bang model
(with vanishing potential). With the same assumptions, when an
 exponential potential for the 
dilaton is included, a scale invariant spectrum ($n=1$) of
adiabatic scalar perturbations is produced under very generic
matching conditions, both in a modified pre-big bang and ekpyrotic
scenario. We also derive the resulting spectrum for arbitrary power
 law scale factors matched to a radiation dominated era.
 \end{abstract}

%\maketitle

\date{\today}
]
\renewcommand{\thefootnote}{\arabic{footnote}} \setcounter{footnote}{0}

\section{Introduction}
Observational cosmology has made enormous progress during the last
couple of years. Most observations seem to agree with the fact
that the total energy density of the universe $\rho$ is very close
to its critical value $\rho_{c}$, $\Om \equiv \rho/\rho_c  =1$,
and it is distributed in the form of pressureless dark matter
$\rho_m$ and dark energy with negative pressure, 
$P_\Lambda \lsim -0.6\rho_\Lambda$,
$\Om=\Om_\La +\Om_m =1$ with $\Om_\La \simeq 0.7$ and $\Om_m
\simeq 0.3$. The clustering properties of the observed universe
agree with a scale invariant spectrum of adiabatic scalar
perturbations, $n\simeq 1$, with or without a tensor component.
Many recent cosmological experiments measure one or several of
these parameters. Most notably cosmic microwave background
anisotropy experiments~\cite{boom,max,cbi}, supernovae type Ia
measurements~\cite{perl,sch}, cluster abundances~\cite{neta},
analysis of the observed galaxy distribution~\cite{2df,sdss}, and
of peculiar velocities~\cite{roman} (see also~\cite{all}).

Although the presence of dark energy, $\Om_\La \neq 0$, remains
very mysterious, inflation explains why $\Om=1$ and $n\simeq 1$.

The basic idea of inflation is simple: If the energy density in a
sufficiently smooth patch of space  is dominated by the potential
energy of some slowly varying scalar field, this patch will expand
very rapidly and evolve into a large, very homogeneous, isotropic
and flat universe. During this rapid expansion, the causal horizon
becomes much larger than the Hubble horizon, alleviating the
horizon problem. In addition, quantum fluctuations in the scalar
field get amplified and grow larger than the Hubble scale,
$H^{-1}$. They then `freeze in' as classical fluctuations in the
energy density or, equivalently, in the geometry, which obey a
scale invariant spectrum.

This standard picture of inflation  does not emerge in a  direct
way from any modern high energy physics model. This makes it very
flexible which is probably one of the main reasons why the basic
picture has survived for so long. If a given model does not work, one
is free to slightly change the potential or other couplings of the
scalar field. This has lead to many different models of inflation
presented in the literature \cite{inflation}. This flexibility may
be considered either as a strong point or as a drawback. It is in
any case certainly very important to investigate whether there are
alternative explanations of the size and the flatness of the
universe and of the observed scale invariant spectrum of adiabatic
scalar fluctuations in the context of modern high energy physics.

In this paper we discuss two attempts in this direction which are both
motivated by string theory: the pre-big bang model~\cite{Ve1,VeGa} and
the ekpyrotic model~\cite{ekp1,ekp2,ekp3}. Even though the
high energy pictures of these models are very different,
the four dimensional low energy effective actions agree and the models
predict the same cosmology at low energy up to possible high energy
'relics'. In the following we call a model of the universe a 'pre-big
bang model' if it contains a low curvature phase before the big bang.
In this sense also the ekpyrotic scenario is a pre-big bang model.

The original pre-big bang model consists just of the dilaton
and the metric, the two low energy degrees of freedom which
are present in every string theory. The presence of the dilaton leads to
a new symmetry called 'scale factor duality' of cosmological solutions: To
each solution for the scale factor $a(t)$ corresponds a solution  $a(t)^{-1}$,
or $a(-t)^{-1}$ if combined with time reversal symmetry. If  $a(t)$ is
an expanding, decelerating solution,  $a(-t)^{-1} \equiv \hat a(\hat t)$ is
an expanding accelerating solution, since
\be
{d\hat a\over d\hat t} =\frac{1}{a^2} { d a \over dt} >0,
\ee
and 
\be
{d^2\hat a\over d{\hat t}^2} = -\frac{1}{a^2} {d^2a\over dt^2} 
+\frac{2}{a^3}\left({d a \over
  dt}\right)^2 >0. 
\ee
The Hubble parameter $\hat H$ of
this 'super-inflating' solution~\cite{Ve1,VeGa}
grows as $\hat t= -t$ increases. The solution approaches trivial flat
space and vanishing couplings in the past, $\hat t \ra -\infty$, and a
curvature singularity in the future,  $\hat t \to0^-$.

In this pre-big bang model, one supposes that curvature and strong coupling
corrections of string theory 'bend' the evolution away from this singularity
into an expanding, decelerating radiation dominated Friedmann model. Several
studies of toy models where this can be achieved have been presented in the
literature (see~\cite{cyril1,ramy,mau,stef}), but they usually just
represent  second order corrections to
the curvature and the coupling, and not full string theory solutions.

It has been shown~\cite{Bru} that a pure dilaton without potential cannot
lead to a scale invariant spectrum of adiabatic scalar fluctuations.
For this reason it has been proposed that fluctuations may be
induced by axions via the so called seed mechanism \cite{axions}.
Axions naturally display a scale invariant
spectrum. However,  the axion seed perturbations are of isocurvature
nature, which is not in agreement with present observations.
Mechanisms which may  convert the axionic isocurvature
fluctuations into adiabatic ones are currently under
investigation~\cite{curvaton}.

In this paper, we will instead repeat the basic arguments of~\cite{Bru},
but we will show that the spectrum of perturbations
which one obtains in the  radiation dominated post-big bang phase
has the spectral index
$n=0$ and not $n=4$ as claimed in~\cite{Bru}. We shall also show  that when
adding an exponential potential to this action, one 
obtains a scale invariant spectrum, $n=1$.
\vspace{2mm}

The high energy picture behind the ekpyrotic scenario, the second
pre-big bang model discussed in this paper, is quite different.
There one starts with a five-dimensional universe containing two
perfectly parallel 3-branes at rest~\cite{ekp1,ekp2}, in a
BPS state. One then supposes
that the two branes approach each other with some very small
initial velocity. It is argued that, from the four-dimensional
point of view of an observer on one of the branes, this situation
corresponds to a collapsing Friedmann universe with a scalar field,
which is related to the distance between the two branes before the
collision. After the collision the solution is supposed to turn into a
radiation dominated Friedmann~\cite{ekp1,ekp2}
(see~\cite{Linde,Lang,Ras} for critics). 

It is assumed that the scalar field is minimally coupled
and has a negative exponential potential $V$ which describes the
attraction of the two branes. The scalar field potential is due to
non-perturbative string corrections but has not been derived from
any string theory, so far. In Refs.~\cite{ekp2,ekp3} it has been
argued that, if $V= -V_0\exp(-c\varphi)$ at low curvature, with
$c\gg 1$, a scale invariant spectrum of scalar perturbations
develops. This result has been criticized in
Refs.~\cite{david1,robi,hwang,hwang2,patrickmartin,tsu}, 
where a spectral index
$n=3$ has been obtained. We shall show here that, even if the
detailed arguments put forward in Refs.~\cite{ekp2,ekp3} might not
be valid, under quite generic (although non trivial) assumptions 
one does obtain the spectral index $n=1$.

Like the original pre-big bang, this
model starts out at low curvature and develops a singularity in the future.
Like there, the belief is that string theory corrections will change the
behavior of the scale factor and the scalar field away from this singular
evolution. In the five dimensional picture, this apparent 'singularity'
corresponds to the collision of the two branes which then should result in the
production of radiation leading to a thermal, radiation dominated Friedmann
model.  We call the phase before the high curvature
regime the 'pre-big bang phase' and the regime after the big bang the
'post-big bang'.

Even if the string theory corrections, which must become important close
to the singularity, are not fully understood, these models are
promising candidates for alternatives to inflation:
They certainly do not suffer from a horizon problem since their age can be
arbitrarily large and is not related to the Hubble time.  They do not
dynamically imply flatness, but this comes from very natural vacuum (for the
original pre-big bang) or BPS (for the ekpyrotic model)
initial conditions which are posed at low curvature. 
Nevertheless, it is well known  that these models
are not very efficient in smoothing out classical 
inhomogeneities~\cite{alessandra} and 
global anisotropies~\cite{kerstin}, and this may remain 
a problem. In the most recent version of the ekpyrotic
model, a cyclic universe, flatness is also a consequence of a period of
exponential expansion in the previous cycle~\cite{cycle}. 
A quite fair
comparison of the ekpyrotic scenario and ordinary inflation is given
in Ref.~\cite{lindenew}.

In this paper we do not address the important debate of the flatness
problem, but we investigate the spectrum of perturbations generated
during the pre-big bang phase. The aim of this paper is to learn as much
as possible about such models without specifying the details of the high
energy phase.

In the next section we write down the modified pre-big bang action and the
action of the ekpyrotic model. We show that they are related by a conformal
transformation and we solve the equations of motion in both Einstein
and string frame. In Sections~III and IV, which are the heart of this
paper, we discuss scalar perturbations and the matching conditions between a
contracting, scalar field dominated phase and an expanding, 
radiation dominated phase. In particular we show that, under certain
well defined conditions, without knowing the details of the matching,
one expects $n=1$ for the modified pre-big bang and the ekpyrotic
model. In Section~V we generalize our  results
to arbitrary power law scale factors matched
to a radiation dominated era.
We end with our conclusions and an outlook.

\section{The background}

The low energy effective action of the original pre-big bang model
is simply gravity with a dilaton $\phi$. Here we modify it by
allowing for a dilaton potential. We assume that we have a
four-dimensional effective theory, any extra dimensions being
frozen at a very small scale. The low energy action for this
theory is therefore~\cite{leesa}
\be
\hat S = \frac{1}{2 \kappa^2} \int d x^4 \sqrt{-\hat g} e^{ -
\phi} \left[ \hat R +(\hat \nabla \phi )^2  - 2\hat V( \phi)
\right], \label{stringframe}
\ee
with $\kappa^2=8\pi G=1/M_P^2$, where $M_P=2.4\times 10^{18}$GeV is
the reduced Planck mass. 
This action is written in the so-called string frame. The hat $\hat{ }$
indicates  that the corresponding
quantities have to be computed using the metric in  this frame.
Therefore $\hat g$, $\hat R$, $\hat\nabla$, and $\hat V$ are
the determinant of the metric, the Riemann scalar, the covariant
derivative, and the dilaton potential, respectively, in the string frame. 
With this action $\phi$ is dimensionless and the usual scalar
field with dimension of mass is simply $M_P \phi$.
Correspondingly, the potential $\hat V$ has dimensions of
(energy)$^2$ and the usual potential is $M_P^2\hat V$. We
 use the metric signature $-+++$.

It is possible to rewrite the action in Eq.~(\ref{stringframe}) in
a conformally related (and physically equivalent) frame. If we 
perform a conformal transformation $g_{\alpha \beta} =\Om^2
\hat g_{\alpha\beta}$ the action is modified to
\begin{eqnarray}
S &= & \frac{1}{2\kappa^2}\int dx^4 \sqrt{-g}
  \Om^{-2}e^{ - \phi}\left[
R + (\nabla \phi)^2 +
\right.
\nonumber \\
&&\left.
+ 6(\nabla \ln \Om)^2 +
6(\nabla \phi \cdot \nabla \ln \Om )
        - 2\Om^{-2} \hat V(\phi) \right].
\end{eqnarray}
When choosing $\Om =\exp(- \phi/2)$,
we can obtain the Einstein frame action,
\begin{equation}
S_{\EI} = \frac{1}{2 \kappa^2} \int d x^4 \sqrt{-g} \left[ R -
\frac{1}{2} (\nabla \phi)^2 - 2 V(\phi) \right],
\label{Einsteinaction}
\end{equation}
where
\be
g_{\alpha \beta} =e^{-\phi} \hat g_{\alpha\beta}  \ \ \mbox{and}
\ \ \ V (\phi) =e^{\phi}\hat V(\phi)
\ee
are the metric and the scalar field potential, respectively,
in the Einstein frame. Eq.~(\ref{Einsteinaction}) is
the action for a minimally coupled scalar field.
Notice that the
dilaton has not been changed by the conformal transformation.
We can also allow for a rescaling of the scalar field,
$\varphi = \phi/\beta$, so that
\be
S_{\EI} = \frac{1}{2 \kappa^2} \int d x^4 \sqrt{-g} \left[ R -
        \frac{1}{2}\beta^2 (\nabla \varphi)^2 -
2 V(\varphi) \right].
\ee

String cosmology and, in particular,
 the original pre-big bang scenario,
has been developed based on  action~(\ref{stringframe}) with  the
dilaton potential set to zero. In our modified pre-big bang model
we will allow a non-zero potential. Since we want to obtain here
the usual scalar field action presented in~\cite{ekp1} starting
from the string cosmology action~(\ref{stringframe}), we have to
require $\beta^2/2   = 1$. This fixes $\beta = \pm \sqrt{2}$. In
terms of the new field $\varphi$ the Einstein frame action now
becomes
\begin{equation}
S_{\EI} = \frac{1}{2 \kappa^2} \int d x^4 \sqrt{-g} \left[ R -
(\nabla \varphi)^2 - 2 V(\varphi) \right] . \label{preaction}
\end{equation}
For an exponential potential
\be
\hat V(\phi)= e^{-\phi} V(\phi) = -V_0 e^{\lambda  \phi},
\label{pbbpotential}
\ee
where $~ \la = -(1+c/\beta)$ with $c\gg 1$,
or equivalently for
\be
V (\varphi) = -V_0 e^{-c \varphi},
\label{ekpypotential}
\ee
we obtain precisely the
low energy effective action of the ekpyrotic scenario~\cite{ekp2,ekp3}.
The interpretation of the field $\varphi$
is however quite different. There $\varphi$ is related to the brane
separation~\cite{ekp2}. At early times when the two branes are
separated by a large distance, the scalar field $\varphi$ is very
big and positive, $\varphi \to \infty$. Therefore the relation between the
string cosmology dilaton  $\phi$ which tends to $-\infty$ for very
early times, $t\rightarrow -\infty$, and the field $\varphi$ of the ekpyrotic
scenario is $\phi= - \sqrt{2} \varphi$, $\beta=-\sqrt{2}$.
Since $c \gg 1$ and $\beta$ is negative, $\lambda > 0$ so that
the potential~(\ref{pbbpotential})
goes asymptotically to zero for very negative dilaton (at early time), 
and does not spoil the initial
conditions of the pre-big bang.

Varying Eq.~(\ref{preaction})   with respect to $\varphi$  we obtain
the equation of motion
\begin{equation}
\Box \varphi - V(\varphi)_{,\varphi}=
\Box\varphi - c V_0 e^{-c\varphi} = 0,
\label{motion1}
\end{equation}
where $\Box= \nabla_\alpha \nabla^\alpha$.
Varying the action with respect
to the metric yields the Einstein equations,
\be
G_{\alpha \beta}=\kappa^2 T_{\alpha\beta} ,
\label{motion2}
\ee
where $T_{\alpha \beta}$ is the energy-momentum tensor of the
scalar field,
\be
\kappa^2 T_{\alpha\beta} =
\nabla_\alpha \varphi  \nabla_\beta\varphi-
    {1\over 2}g_{\alpha\beta} \left[  (\nabla \varphi)^2
+2 V (\varphi) \right].
\ee

We want to consider
a flat homogeneous and isotropic universe with metric
$ds^2 =-d t^2+ a^2dx^2$. In this case
Eq.~(\ref{motion1}) becomes
\be
\ddot\varphi + 3 H \dot \varphi + V_{, \varphi}=0,
\label{ekpy2} \ee
where the over-dot is a derivative with respect to
the cosmic time $t$, and (\ref{motion2}) turns into the Friedmann equation,
\begin{equation}
H^2=\frac{\kappa^2}{3} \rho = \frac{1}{6} \dot \varphi^2 + \frac{1}{3}
V(\varphi). \label{ekpy1}
\end{equation}
Eqs.~(\ref{ekpy2},\ref{ekpy1}) have the `ekpyrotic solution'~\cite{ekp2}
\be
a(t) = (- t)^p
,~~~~~  \varphi (t) = {2\over c}\ln(- M t),
\ee
with
\be
p=\frac{2}{c^2},~~~~~ M^2={ V_0\over p(1-3p)}.
\ee
At first it may seem strange that the enthalpy $w \equiv P/\rho$ and
the sound speed $ c_s^2 \equiv \dot P/\dot\rho$ are much larger
than one, $c_s^2 =w \gg 1$, for small values of $p$ (large $c$),
\be
w = {(1/2)\dot\varphi^2 - V\over (1/2)\dot\varphi^2 +V}
 =  c_s^2 = {2\over 3p} -1. \label{definiti}
\ee
On the other hand, as long as we
concentrate on a time
interval  bounded away from the singularity, we can always split
the potential into $V =V_1(\varphi) + V_2$, where $V_2$ is a very
negative constant and $V_1$ is always positive. Interpreting $V_2$ as a
negative cosmological constant, we have
\be
-1<w_1={(1/2)\dot\varphi^2 -V_1\over (1/2)\dot\varphi^2 +V_1}<1,
\ee
as well as $-1<c_1^2<1$ and $w_2=c_2^2=-1$. However, since $\Om_1
 =\rho_1/(\rho_1 +V_2) \gg 1$ and $\Om_2=V_2/(\rho_1 +V_2)\ll -1$, the
'effective' $w =w_1\Om_1 -\Om_2$ can become much larger than $1$ without
implying any pathological or even acausal behavior of the scalar field
'fluid'.

We shall see that the perturbations generated in this collapse phase acquire
a scale invariant spectrum only if the collapse proceeds very slowly,
{\em i.e.}~when
$0<p\ll 1$. In the ekpyrotic scenario the collapse is followed by an
expanding phase. Shortly before the bounce at $t \to 0^-$, when the
scalar field, after having become negative,
goes to minus infinity, $\varphi \ra -\infty$, the shape of the
potential has to change from the exponential expression, and
turn upwards in such a way that $V\ra 0$ for  $\varphi \ra -\infty$.

Let us give here, for completeness, the equations derived 
from the string frame action Eq.~(\ref{stringframe}), 
where the potential $\hat V(\phi)$ is
given by Eq.~(\ref{pbbpotential}), and their
solutions. 
By varying this action with
respect to the field $\phi$ we obtain
\begin{equation}
2\hat\nabla_\alpha \hat \nabla^\alpha 
\phi - (\hat\nabla\phi)^2 +\hat R -2\hat V
+2\hat V_{,\phi} =0 . \label{sphi}
\end{equation}
Varying the action with respect to $\hat g^{\alpha\beta}$ yields
\begin{equation}
\hat G_{\alpha\beta}
= - \hat\nabla_\alpha \hat\nabla_\beta\phi -\frac{1}{2}
\hat g_{\alpha\beta}\left[ (\hat\nabla\phi)^2   -2
\hat\nabla_\alpha \hat \nabla^\alpha \phi +2 \hat V \right] .   \label{sEin}
\end{equation}
For a homogeneous and isotropic
universe with spatially flat sections, Eqs.~(\ref{sphi}) and (\ref{sEin})
reduce to
\begin{eqnarray}
\ddot \phi + 3 \hat H \dot \phi - \dot \phi^2 + 2\hat V +
2\hat V_{, \phi}&=&0,
\label{field}\\
\hat H^2 - \hat H \dot \phi + \frac{1}{6} \dot \phi^2 -
\frac{1}{3} \hat V&=&0, \label{fried}
\end{eqnarray}
where the over-dot here refers to cosmic time in the string
frame, $\hat t$.

To find a solution to these equations we can simply transform
the solution found in the Einstein frame using the relations
\be
d\hat t =e^{\phi/2}d t= e^{-\varphi/\sqrt{2} } d t,~~ \hat a =
e^{\phi/2} a= e^{-\varphi/\sqrt{2} } a . \label{19}
\ee
The first relation gives
\be
 - \hat M \hat t={(- M t)^{1-\sqrt{p}}} ,
\ee
where $\hat M=M(1-\sqrt{p})$. For small $p$, $p\ll 1$, $\hat t$ is
very close to $t$ and, as long as $p<1$, $\hat t$ grows from
$-\infty$ to $0$ with $t$. Inserting the ekpyrotic solutions in
expressions (\ref{19}) for $\hat a $ and $\phi$, we obtain
\be
\hat a = (-\hat M\hat t)^{-\sqrt{p}},
\ee
and
\be
\phi = - \sqrt{2} \varphi = -{2 \sqrt{p} \over 1 - \sqrt{p}}
\ln(-\hat M\hat t),
\ee
up to possible integration constants which we have fixed to obtain
$\hat a=a$ and $\hat t =t$ in the limit $p \ra 0$.

In this section we have first shown that, from a purely
four-dimensional point of view the ekpyrotic scenario is
equivalent to the pre-big bang scenario when the dilaton has an
exponential potential that tends to zero at small coupling. In doing so
we have presented the equations for these models, written in the
string and Einstein frames, and we have written down the solutions that
hold in either frames. These solutions are useful
for discussing perturbations, which is the subject of the next section.

\section{Scalar Perturbations}

We now want to study linear perturbations of a generic universe
dominated by a minimally coupled scalar field with an exponential
potential or an adiabatic fluid with $w=c_s^2=$ constant. This
last condition is automatically satisfied for a scalar field with
exponential potential.

As discussed in the previous section,
pre-big bang expansion in the string frame is equivalent to contraction
in the Einstein frame, where the dilaton is minimally coupled.
Therefore, pre-big bang with a dilaton corresponds to a
collapsing universe dominated by a minimally coupled scalar field and
is included in our study.
It is important to note that
physical quantities, like the spectral index or the perturbation amplitude
are frame independent but they are more easily computed
in the Einstein frame
where linear perturbation theory
is well established (see,
{\em e.g.}~the reviews~\cite{slava,ruth}).

To discuss perturbations we work mainly in conformal time $\eta$, which is
related to the physical time $t$ by $a d\eta =dt$. The derivative 
with respect to
conformal time is denoted by a prime, $'$.
For the sake of simplicity we neglect a possible curvature of
the spatial sections.
In a flat  universe dominated by a fluid or a scalar field
with energy density $\rho$ and
pressure $P$ the background Friedmann equations are
\bea
\HH^2 &=& \frac{\kappa^2}{3} \rho a^2, \label{frieee1}\\
\HH' &=& - {\kappa^2 \over 6} (\rho+3 P) a^2 = - \HH^2 
\frac{1+3 w}{2}   \label{frieee2},
\eea
where $\HH = a'/a$.

If the energy density is dominated by a scalar field,
we have
\bea
\kappa^2 \rho &=& \frac{1}{2a^2} {\varphi'}^2 + V(\varphi), \\
\kappa^2 P &=& \frac{1}{2a^2} {\varphi'}^2 - V(\varphi),
\eea
and
\be
w +1= \frac{{\varphi'}^2}{3 \HH^2} .
\ee
When $w=c_s^2=$ constant, the solution to the Friedmann equation
is a power law. In terms of conformal time $\eta$ it is given by
\be
a=\left| \frac{\eta}{\eta_1} \right|^q,~~~ q= {2\over 1+3w},~~~
\HH = {q\over \eta}
,~~~ \HH' = -{q\over\eta^{2}},
\label{solus}
\ee
where we have chosen the normalization constant $\eta_1$ such that
 $-\eta_1 < 0$ is a very small negative time at which (higher order)
corrections to the scalar field action become important. Since
$a(\eta_1)=1$, $\eta_1=a(\eta_1)\eta_1\sim t_1$ corresponds to a 
physical quantity, {\em e.g.}\ the string scale in the pre-big bang model,
$1/\eta_1\sim 10^{17}$ GeV.
Comparing Eq.~(\ref{solus})
with the ekpyrotic solutions in terms of physical time, we find $q=p/(1-p)$.

Let us now perturb the metric.  In longitudinal gauge and in
absence of anisotropic stresses, as it is the case for perfect
fluids and for scalar fields, scalar metric perturbations are
given by
\be
 ds^2 = a^2(\eta)[-(1+ 2\Psi)d\eta^2 + (1- 2\Psi)\delta_{ij}dx^idx^j].
\label{longi}
\ee
In this gauge, the metric perturbation $\Psi$ corresponds to the
gauge invariant Bardeen potential. Without gauge fixing the latter
is given by a more complicated expressions of the metric
perturbations~\cite{slava,ruth,Ba}. The scalar field $\varphi$ is
also   perturbed so that it can be divided into $\varphi(\eta)$
satisfying the background equation (\ref{ekpy2}), and a
perturbation $\delta \varphi(\eta,{\bf x})$.

We now want to compute the spectrum of metric perturbations 
generated from vacuum initial conditions.
Generically, $\Psi$  satisfies the equation~\cite{slava,ruth}
\bea
\Psi'' +  3 \HH (1+c_s^2) \Psi' + \qquad\qquad  && \nonumber \\
\qquad\qquad ( 2 \HH' + (1+3 c_s^2) \HH^2 - \Upsilon \Delta ) \Psi
& = & 0.
\label{II}
\eea
For adiabatic perturbations of a fluid, one finds  $\Upsilon = c_s^2$,
where $c_s^2$ is the adiabatic sound speed, while for  a simple scalar field
one finds $\Upsilon = 1$ (see, {\em e.g.}\ Ref.~\cite{slava}). Hence for
a non-vanishing potential, $V\neq 0$ and hence $c_s^2\neq 1$, simple
scalar field perturbations are not adiabatic in a thermodynamic sense.

If we restrict ourself to the case, $w=c_s^2=$ constant, the mass
term in Eq.~({\ref{II}),  $2 \HH' + (1+3 c_s^2) \HH^2$, vanishes by the use of the background
Einstein equations, Eqs.~(\ref{frieee1}) and (\ref{frieee2}).
Thus, for scalar perturbations we obtain nearly the same equation as for
tensor perturbations, which we can write in terms of Fourier modes
as
\be
\Psi'' +  3\HH(1+w)\Psi' +
\Upsilon k^2 \Psi = 0  \label{Phi''}.
\ee
This equation is valid in both phases of the universe,
before and after the big bang, depending on the
corresponding value of $w$ and $\Upsilon$.
We call $\Psi_-$ the solutions obtained in the pre-big bang collapsing phase
and $\Psi_+$ the one obtained in the radiation dominated phase.
In the following we will work in Fourier space.

Let us now define the variable $u$
in order to simplify Eq.~(\ref{Phi''}) \cite{slava}. We set
\be
u = \frac{M_P}{\HH} a \Psi.
\ee
Eq.~(\ref{Phi''}) can then be written in terms of $u$ as
\be
u'' +\left(\Upsilon k^2 - a (1/a)'' \right)
u =0 \label{u''}.
\ee

Let us now suppose that the collapsing (or pre-big bang)
phase $\eta<-\eta_1$ is dominated by the scalar field so that $\Upsilon=1$.
Eq.~(\ref{u''}) then has the general solution
\be
u = (k|\eta|)^{\frac{1}{2}} [ C(k) H_\mu^{(1)}(k\eta) +
        D(k) H_\mu^{(2)}(k\eta) ]  \label{solu},
\ee
with $\mu =q+1/2$. Here $H_\mu^{(i)}$ is the Hankel function of the
$i$-th kind and of order $\mu$. One can generalize this solution to the case
of a fluid dominated universe simply by replacing $k\eta$ by $c_sk\eta$.
This solution has to be generated from the incoming vacuum,
so we assume that, for $  k | \eta| \gg 1 $,
\be
\lim_{  \eta \to - \infty} u = \frac{e^{-ik\eta}}
{k^{3/2}}.
\ee
This assumption corresponds  to normalizing  the canonical
variable which diagonalizes the perturbed second order action
(called $v$ in \cite{slava}) or equivalently the perturbation of
the scalar field, $\delta \varphi$, to quantum vacuum
fluctuations. With this normalization,  the $H^{(1)}_\mu$ mode,
which approaches $\exp(ik\eta)$ for $ k | \eta| \gg 1 $, has to be
absent, $C(k)=0$, and the solution to Eq.~(\ref{Phi''}) becomes
\be
\Psi_- (k, \eta) =  { q \over M_{P}a \eta}
        D(k)(k|\eta|)^{1/2} H_\mu^{(2)}(k\eta) \label{solPsi},
\ee
where
\be
D(k) = \sqrt{ \pi/2} k^{-3/2},
\ee
modulo some irrelevant phase.

At late time
$k |\eta|\ll 1$, this solution approaches
\be
\Psi_-(k,\eta) \simeq  A_{-}(k){\HH\over a^2} +B_-(k),
        \label{Psi-} \label{solls}
\ee
where $A_-$ and $B_-$ are determined by the exact solution (\ref{solPsi})
(up to logarithmic corrections),
\bea
A_-(k) &\simeq&
\frac{2^\mu \Ga(\mu)}{M_P \eta_1^q } k^{-\mu-1},
\label{A-}\\
B_-(k) &\simeq&
\frac{\eta_1^q }{ M_P 2^\mu \Ga(\mu+1)} k^{\mu-1}~
. \label{B-}
\eea
The result~(\ref{Psi-}) can be found directly by solving Eq.~(\ref{Phi''})
neglecting the $k^2$-term. The full solution is however needed to
determine the pre-factors $A_-(k)$ and $B_-(k)$ from the vacuum
initial condition. The $A_-$-mode grows during the pre-big bang phase
and becomes much larger than the constant $B_-$-mode.

In the original pre-big bang, where the dilaton has no potential,
{\em i.e.}~$w=c_s^2=1$ and hence $q=1/2$, we have $\mu=1$. The
$A_-$-mode then has an $n=0$ spectrum, $|A_-|^2k^3\propto
k^{-1}\propto k^{n-1}$, while the $B_-$-mode corresponds to $n=4$,
$|B_-|^2k^3\propto k^{3}\propto k^{n-1}$. If we have an exponential
potential as for the ekpyrotic model such that $p \ll 1$, and therefore $q\ll
1$, we have $\mu \simeq 1/2$ and hence $|A_-|^2k^3$ is
$k$-independent. The $A_-$-mode has a scale invariant
 spectrum, $n=1$, while $|B_-|^2k^3\propto k^2$, which corresponds to a blue
spectrum, $n=3$.

If the $A_-$-mode has a red spectrum, as in the original pre-big bang
scenario, we need to discuss  its amplitude on large scales. 
It has been
shown in~\cite{Bru} that a red ($n=0$) $A_-$-mode does not invalidate linear
perturbation theory during the pre-big bang phase. Geometrically
meaningful quantities like
$C_{\alpha\beta\gamma\delta}C^{\alpha\beta\gamma\delta}/R^2 \equiv
\Delta^2 $, where $C_{\mu \nu \gamma \delta}$ is the Weyl tensor and
$R$ is the curvature scalar, remain small. In fact $\Delta^2
\propto  |(k\eta)^2 \Psi|^2k^3$.
%or the projection of the density
%gradient onto the surfaces normal to $\nabla_\alpha\phi$ 
We can therefore continue to use the Bardeen potential
even though it may become large for certain $k$-modes. 
However, a red spectrum leads to serious problems in the subsequent radiation
era where the Bardeen potential is constant on super horizon scales
and $\Delta^2$ grows larger than unity at horizon entry, $k\eta \sim
1$, for large scales. 

In the modified pre-big bang models discussed here, 
this problem does not occur,
since $A_-$ has a scale invariant spectrum.

At very early time {\em after} the big bang, in the radiation dominated phase,
we can neglect the term $\Upsilon k^2 =  k^2/3$  in Eq.~(\ref{u''}).
We then have the same type of solution for super horizon modes,
\be
\Psi_+(k,\eta) =  A_+ (k){\HH\over a^2} +B_+ (k). \label{sollr}
\ee
In the next section we will work out the matching conditions
between this solution and Eq.~(\ref{solls}), in order to
determine the coefficients $A_+$ and $B_+$.

\section{Matching conditions}

We suppose that the solution given in Eq.~(\ref{solls}) holds until
$\eta=-\eta_1$, where higher order corrections begin to play a role. These
corrections  may be quite different for the modified pre-big bang
model and for the ekpyrotic model, but in both cases they are
supposed to lead over to a radiation dominated Friedmann model.
Here we do not want to argue about the nature of the corrections
and how to determine them from string theory (even if this
probably has to be considered as the most difficult and the main
problem of these models), but we study which statements can be
made under certain assumptions on the transition. For this we
neglect the details of the transition and match our pre-big bang
solution at $\eta=-\eta_1$ to a radiation dominated universe at
$\eta=+\eta_1$. In other words we suppose that the slice of
spacetime `squeezed' between $-\eta_1$ and $\eta_1$ is so thin
compared to the scales we are interested in, that it can be
replaced by a spacelike hypersurface. Therefore we can
consistently use the thin shell formalism and apply the Israel
junction conditions~\cite{isi} for surface layers on the $\eta =
\pm \eta_1$ hypersurface, in order to match the spacetime manifold
${\cal M}_-$ before the big bang to the spacetime manifold ${\cal
M}_+$ after.

\subsection{Matching the background}

Before specifying the matching of the perturbations, we have to
match the backgrounds, {\em i.e.}\ we have to impose the Israel junction
conditions on the scale factor $a$ and its first derivative. These
conditions require the continuity of the induced metric,
\be
q_{\alpha \beta}=g_{\alpha \beta} +n_\alpha n_\beta,
\ee
where $n_\alpha$ is the normal to the $\eta=$ constant
hypersurface, on the matching hypersurface $\eta=\pm\eta_1$. Thus
we have
\be
[q_{\alpha \beta}]_{\pm} =0,
\label{i1}
\ee
where we define
\be
[h]_{\pm} \equiv   \lim_{\eta \searrow\eta_1} \left( h(\eta)
- h(-\eta) \right) \equiv h_+ - h_- ,
\label{limit}
\ee
for an arbitrary function $h(\eta)$. Here $\eta \searrow\eta_1$
indicates the right hand limit, {\em i.e.}\ $\eta$ is decreasing towards
$\eta_1$. 

Our conformal time coordinate $\eta$ itself
jumps,
\be
[\eta]_\pm = 2\eta_1.
\ee
This simply means that the coordinates of ${\cal M}_-$ and ${\cal
M}_+$ are well defined only on the intervals 
$ \eta \in (-\infty, -\eta_1]$ and
$ \eta \in [\eta_1, \infty)$, respectively.
The limit (\ref{limit}) 
is well defined for every function which is
continuous, monotonic and bounded in open intervals $(-\eta_2, -\eta_1)$ and
$(\eta_1,\eta_2)$, with $\eta_2>\eta_1$, even if their value at
$\pm\eta_1$ is not defined.

Eq.~(\ref{i1}) implies $ a_+ = a_{-}=a_\pm$. According to our
normalization of the scale factor, Eq.~(\ref{solus}),
$a_\pm =1$. We nevertheless
prefer to leave $a_\pm$ in all the expressions where it appears,
so that its normalization can be conveniently changed.

The second Israel junction condition concerns the extrinsic curvature
$K_\mu^\nu$ on the matching hypersurface with normal $n^\al$,
\be
K_{\alpha\beta}= {1\over 2} ( q_{\alpha}^{ \ \rho}  
\nabla_{\rho} n_\beta + q_{\beta}^{ \ \rho}
  \nabla_{\rho} n_\alpha).
\ee
In a Friedmann universe this is
\be
  K^i_j = -\left({a'\over a^2}\right)\de^i_j = -
  {{\cal H}\over a}\delta^i_j.
\ee
The derivative $a'$ changes sign in the transition from a
contracting to an expanding phase. Hence, the extrinsic curvature
is discontinuous in the four-dimensional, low energy picture if we simply
'glue' the contracting phase  to the expanding phase with
opposite sign for $a'$ and conformal time $\eta=+\eta_1$. On the
other hand, the Israel junction conditions allow for the
existence of a surface stress tensor,
\be
[K^i_j]_\pm=\kappa^2 S^i_j, \label{i2}
\ee
which in our case is non vanishing and diagonal, and it is
characterized by a negative surface tension $P_s <0 $,
\be
[K^i_j]_\pm= -\frac{{\cal H}_+-{\cal H}_-}{a_\pm}\delta^i_j =
\kappa^2 P_s\de^i_j. \label{backjump}
\ee
Within the four dimensional picture we have no explanation for
this surface tension; it has to be introduced by hand in order for
the extrinsic curvature to jump. Eq.~(\ref{backjump}) is a
possibility to 'escape' the violation of the weak energy
condition, $\rho+P < 0$, which is needed for a smooth transition
from collapse to expansion. This has been one of the objections to
the ekpyrotic scenario in Ref.~\cite{linde}. Of course for $\eta
=\eta_1$ the combination $\rho + P + P_s\delta(\eta-\eta_1)$ becomes
negative, which, in the widest sense, can also be interpreted as an
'effective' violation of the weak energy condition. Clearly, this is the
simplest way of connecting a contracting phase to an expanding
phase, but it is relatively close to an approach motivated from
the five-dimensional picture, where the singularity at $a=0$
becomes a narrow 'throat' \cite{ekp2}. Here we replace this throat
by a stiff 'collar' whose length we neglect (see also
\cite{ekp3}).

\subsection{Matching the perturbations}
\label{matchsec}

Let us now perturb the Israel junction conditions (\ref{i1}) and
(\ref{i2}). Instead of considering the $\eta = \eta_1$
hypersurface we want, in general, to consider a hypersurface which is
linearly perturbed from it, defined by
$\tilde\eta = \eta+T = \eta_1$, where $T$ is a small
perturbation. 
The jump is now realized on the perturbed hypersurface $\tilde \eta = \eta_1$,
\be
[h]_{\pm} \equiv   \lim_{\tilde \eta \searrow\eta_1} \left( h(\tilde \eta)
- h(- \tilde \eta) \right) \equiv h_+ - h_- ,
\ee
and in principle we cannot say anything about the
continuity of $T$, which is also
allowed to jump,
\be
[T]_\pm = [ \tilde \eta - \eta]_\pm = 2\eta_1 - [ \eta]_\pm. \label{Tjump}
\ee
Nonetheless, this jump should be always small as it will become
clear below.

We assume that the old coordinates $(\eta, x^i)$ are those of
longitudinal gauge, so that the metric perturbations are given by
Eq.~(\ref{longi}), but we want to determine the perturbation of
the  Israel junction conditions in the coordinate system
$(\tilde\eta, x^i)$ on the surfaces $\tilde\eta=$ constant. The
metric in this coordinate system is given by (see {\em e.g.}
\cite{ruth})
\bea
d \tilde{s}^2 &=& a^2(\tilde\eta)\{-(1+ 2\Psi -2(\HH T
+T'))d\tilde\eta^2  + 2T,_{i}d\tilde\eta dx^i  \nonumber \\ &&
+(1- 2\Psi -2 \HH T)\delta_{ij}dx^idx^j\}. \label{shorti}
\eea
Hence  the perturbation of the normal to the $\tilde \eta=$
constant slices is
\be
\tilde \delta n =\frac{1}{a} \{ (-\Psi +\HH T +T') \dd_{\tilde
\eta} -T,_i\dd_i\} ,
\ee
and the extrinsic curvature is given by \cite{nathalie}
\begin{equation}
\tilde \delta K^i_j=\frac{1}{a}\left\{\Psi'+{\cal H}\Psi +({\cal
H}'-{\cal H}^2)T \right\} \delta^i_j +T^{,i}_{,j} .
\label{m2}
\end{equation}

The matching conditions for the perturbations are obtained by
perturbing Eqs.~(\ref{i1}) and (\ref{i2}) on the $\tilde \eta= \eta_1$
hypersurface. They become
\be
[ \tilde \delta q^i_j ]_\pm=0, \qquad [ \tilde \delta K^i_j]_\pm=
\kappa^2 \tilde \delta S^i_j.
\ee
From the above expressions for $\tilde \delta g_{\mu \nu}$ and
$\tilde \delta n^\mu$, the continuity of the perturbation of the
induced metric $\delta q_{ij}$ on the $\tilde{\eta}= \eta_1$ 
hypersurface leads to
\begin{equation}
\left[\Psi+{\cal H}T\right]_\pm=0.
\label{m1}
\end{equation}

For reasons that become clear below, we
assume in the following
that $T=\tilde \eta -\eta$, the lapse of time  between
the background value $\eta$ and the perturbed value $\tilde \eta$,
remains a small perturbation on large scales. This implies  
that also $[T]_\pm$ has to remain small. What is the meaning
of 'small perturbation' in this context? Once a gauge is fixed,
the Bardeen potential $\Psi$ is the only degree of freedom
characterizing the perturbations. For dimensional reasons, it is
natural to expect $T$ to be given as a linear combination of
$\Psi$ and $\Psi'$, in terms of %  $k\eta$,
\be
T=\eta P(k \eta) \Psi + \eta^2 Q(k \eta) \Psi',
\ee
where $P$ and $Q$ are polynomials of $k \eta$, which may have 
$\eta/\eta_1$ dependent coefficients. Here we assume that these
polynomials do not contain any negative power of $k \eta$, {\em i.e.}
that
\be
|{T / \eta \Psi} | \sim  |{T /\eta^2 \Psi'} | 
\sim  |P(k\eta)| + |Q(k\eta)| \stackrel{ k\to 0}{ \longrightarrow}
 {\rm finite}. \label{assumption1}
\ee
On large scales $T$ grows with scale at most as $\Psi$ or $\Psi'$. 

The reason for this is that we want that the $\tilde \eta= \eta_1$ 
hypersurface does not arbitrarily diverge from the $\eta=\eta_1$
hypersurface on large scales. In other words, we require
the time at which the bounce happens to be stable under large
scale perturbations. 
It is clear that this assumption is not entirely trivial. It limits
somewhat the large scale power of the `new physics' which is needed
to convert contraction into expansion. This new physics may not induce
very strong infrared perturbations, which is  very
reasonable and confirmed by numerical examples on pre-big bang
models \cite{cyril2}.

Under this assumption the anisotropic term on the right hand side
of Eq.~(\ref{m2}), $\partial^i
\partial_j T$, is negligible on large scales and we shall not discuss
the possible, but sub-dominant, anisotropic surface stresses in what
follows. On super horizon scales the perturbation of the extrinsic
curvature is dominated by the trace part,
$\tilde \delta K^i_j = (\tilde \delta K)\delta^i_j$ with
\begin{equation}
\tilde \de K = \frac{1}{a}\left\{\Psi'+{\cal H}\Psi+({\cal
H}'-{\cal H}^2)T \right\}. \label{deltaK}
\end{equation}
The matching conditions for the perturbations become
Eqs.~(\ref{m1}), and
\begin{equation}\label{m3}
\left[\Psi'+{\cal H}\Psi+({\cal H}'-{\cal H}^2)T\right]_\pm
  =\kappa^2 a_\pm \tilde \delta P_{s},
\end{equation}
where $\tilde \de P_{s}$ is the perturbation of the surface
tension.

The condition posed in Eq.~(\ref{assumption1}) has the following
important consequences: from Eq.~(\ref{deltaK}) we see that with $T$
not being 'redder' than $\Psi$ and $\Psi'$, also
$\tilde \delta K$ has typically the same $k$-dependence as $\Psi$ or
$\Psi'$. Therefore it remains small (of the same order as $\Psi$
or $\Psi'$ in $k$) when $k \eta_1$ tends to $0$,
\be
\tilde \delta K /(\HH \Psi), ~ \tilde \delta K /\Psi' ~ 
\stackrel{ k\to 0}{ \longrightarrow} ~ {\rm finite}. \label{assumption2}
\ee
From Eq.~(\ref{m3}) we then infer that $\tilde \delta P_s$ may as well
have a non-trivial $k$-behavior but it remains small on large
scales,
\be
\tilde \delta P_s /({\cal H}\Psi), ~ \tilde \delta P_s /\Psi' ~ 
\stackrel{ k\to 0}{ \longrightarrow} ~
{\rm finite}. \label{assumption3}
\ee
The $k$-dependence of $\tilde \delta P_s$ may become important
when matching the perturbations but it cannot dominate on large scales.

The assumptions (\ref{assumption1})  and its consequences
(\ref{assumption2}) and (\ref{assumption3}) become important
in Sec.~\ref{ububu} where we try to derive a general result from
these matching conditions. First, let us discuss some examples.

\subsection{Two examples}

The  matching conditions (\ref{m1}) and (\ref{m3}), which the
unknown details of the transition have to determine, fix the
coefficients $A_+(k)$ and $B_+(k)$. So far, in the literature, for
inflation~\cite{nathalie} as well as for the
ekpyrotic scenario~\cite{david1,robi,hwang,hwang2,robi2}, 
the hypersurface
on which the matching has been performed was always chosen to be
the constant energy hypersurface, $\rho + \delta \rho=$ constant.
In this case, $T=\de\rho/\rho'$. 

The perturbed Einstein equations
give (see {\em e.g.}\ \cite{ruth}, Eqs.~(2.45) and  (2.46), and use
$\de\rho= \rho D_s$ in longitudinal gauge),
\bea
{\de\rho\over \rho} &=& -{2\over {\cal H}^2}\left\{(3k^2+{\cal
H}^2)\Psi +{\cal H}\Psi' \right\}  \nonumber \\
 &\simeq&  - 2\left(\Psi + {\cal H}^{-1}\Psi' \right),
\eea
on super horizon scales. With $\rho' =  2\rho(\HH'-\HH^2)/\HH $ we
have
\be
T = \de \rho/\rho' \simeq {-1\over \HH'-\HH^2}(\HH\Psi
  +\Psi'). \label{mani2}
\ee
Eq.~(\ref{m1}) then leads to
\begin{equation}\label{me1}
\left[ \Psi-  {\HH\over \HH'-\HH^2}(\HH \Psi +\Psi')\right]_\pm
\equiv [\zeta]_\pm = 0,
\end{equation}
where $\zeta$ is the curvature perturbation introduced by
Bardeen~\cite{Ba}. Furthermore, using Eq.~(\ref{mani2}), one finds
 that $\tilde \delta K_{ij}=0$ on large scales and we obtain $[\tilde \de
K]_\pm \equiv 0$. Hence, this matching condition can be satisfied
only if the surface tension $P_s$ is unperturbed, $\tilde \de
P_s\equiv 0$.

These matching conditions are often used in inflationary models to
go from the inflationary phase to the Friedmann radiation
dominated phase. The difference with inflationary models is that
here $\HH$ jumps.  Furthermore,  $\Psi$ in general will not be
continuous at the transition, since even if $T$ is continuous, $\HH T$ is
not. Notice that, even though $\HH$ jumps at the transition from contraction
to expansion, and hence $\HH'$ contains a Dirac delta-function,
$[T]_\pm$ is well defined as it is a continuous, bounded, monotonic
function in some open intervals
$(-\eta_2,-\eta_1)$ and $(\eta_1,\eta_2)$.

Inserting ansatz~(\ref{solls}) and (\ref{sollr}) in the continuity
condition for the metric, Eq.~(\ref{me1}), yields
\be
B_+\left( \HH'_+ -2\HH_+^2\over\HH'_+ - \HH_+^2 \right) =
B_-\left(\HH'_+ -2\HH_-^2\over\HH'_- - \HH_-^2 \right) .
\ee
Clearly, since $B_+$ couples only to $B_-$ it inherits the blue
spectrum of $B_-$. This is the main argument of
Refs.~\cite{david1,robi,hwang,hwang2,patrickmartin,tsu} 
against the ekpyrotic model. As
we shall see below, this is also the
matching condition which leads to the $n=4$ spectrum in the pre-big bang
model given in Ref.~\cite{Bru}.

There are two subtleties which have been left out in this
argument. The first one is obvious: the surface tension $P_s$, the
only ingredient of the high energy theory in this approach, may
well also have a perturbation $\tilde \de P_s$, requiring $[\tilde \de K]_\pm =
\kappa^2 \tilde \de P_s \neq 0$. If this is the case, the matching cannot be
defined on the constant energy hypersurfaces,
$T=\de\rho/\rho'$. Secondly, and more importantly, in this 
model where contraction goes over to expansion, a transition
surface with a physical surface tension is required and this
surface does need not to agree with the $\rho+\delta \rho=$
constant!

As a concrete example, let us simply assume that this matching surface is
given by the condition that its shear vanishes. This is  actually
just the $\eta=$ constant surface in longitudinal gauge, hence we
have $T=0$ in Eqs.~(\ref{m1}) and (\ref{m3}). The junction conditions on
super horizon scales then become
\bea
[\Psi]_\pm  &= & 0, \label{m0s1} \\
\left[ \HH\Psi + \Psi' \right]_\pm &=& a_\pm \ka^2\de P_s. \label{ms02}
\eea
For our general solutions (\ref{solls}) and (\ref{sollr}) this gives
\bea
A_+ &=& {\HH_-\over \HH_+}A_-  +
{ a^2_\pm \over \HH_+}(B_- -B_+) \label{A+s0}\\
 B_+ &=& \left({\HH_+(\HH_-'/\HH_- -\HH_-) - \HH_+'+\HH_+^2
        \over 2\HH_+^2 -\HH_+'}\right) {\HH_-\over a^2_\pm} A_-
 \nonumber  \\ &&  +
\left( 1 + {\HH_-\HH_+ - \HH_+^2\over 2\HH_+^2-\HH_+'} \right)  B_-
 \nonumber \\ && +
{\HH_+\over 2\HH_+^2-\HH_+'}  \ka^2a_\pm\de P_s  .  \label{B+s0}
\eea

Alternatively, we can express the matching conditions in terms of $\zeta$
given in Eq.~(\ref{me1})
and its canonically conjugate variable $\Pi$ defined in Ref.~\cite{gabi},
by
\be
\Pi = 2 M_P^2 k^2 \frac{a^2}{\HH} \Psi .
\label{Pi}
\ee
On super horizons scales we have
\bea
\zeta &=& \left(1- \frac{\HH^2}{\HH' - \HH^2} \right) B(k), \\
\Pi &=& 2 M_P^2 k^2 \left(A(k) + \frac{a^2}{\HH} B(k)\right) .
%\to 2 M_P^2 A(k) k^2.
\eea
The perturbation
variable $\zeta$ is constant and proportional
to the constant $B(k)$  while
its conjugate  momentum $\Pi$ is proportional  to $A(k) k^2$ and constant
up to a decaying part proportional to $B(k)$ which will be
negligible at the time $-\eta_1$, when we  impose the matching conditions.

On the zero shear hypersurface we can write the matching
conditions of the perturbations
in terms of
 $\zeta$ and $\Pi$ as
\bea
\left[ \HH \Pi \right]_\pm &=&0, \\
\left[ (\HH' - \HH^2) \left( \frac{\kappa^2}{2 k^2 a^2} \Pi
- \frac{\zeta}{\HH}   \right)  \right]_\pm &=&  a \kappa^2 \delta P_s .
\eea
Therefore we have
\bea
\Pi_+ &=& \frac{\HH_+}{\HH_-} \Pi_-, \\
\zeta_+ &=& \frac{\HH_+}{\HH_-}
    \left(\frac{\HH'_- - \HH^2_-}{\HH_+' - \HH_+^2}\right)
\zeta_- \nonumber \\
&& + \frac{\kappa^2}{2 k^2 a_\pm^2} \left( \HH_-
- \frac{\HH_-' - \HH_-^2}{\HH_+' - \HH_+^2} \HH_+   \right) \Pi_- \nonumber \\
&&- \frac{\HH_+}{\HH_+' - \HH_+^2} a_\pm \kappa^2 \delta P_s.
\eea
Hence, using matching conditions on the zero shear hypersurface,
$\zeta$ acquires, in the radiation dominated era, a mode
$\propto \Pi_- k^{-2} \propto A_-$ which has a spectral index $n=1-2
q$ of $A_-$.
In terms of $A_+$ and $B_+$ this leads again to Eqs.~(\ref{A+s0}) and
(\ref{B+s0}).

As $A_-$ represents the growing mode during the contracting phase,
$|A_-\HH/a^2|$ is much larger than $|B_-|$, and the spectrum of
$B_+$ inherits the scale invariant spectrum of $A_-$. It is easy
to see from  pure sign considerations that the pre-factor of
$A_-$ in Eq.~(\ref{B+s0}) does not vanish.

\subsection{A more general treatment}
\label{ububu}

As we have seen, the important question is to determine the
correct matching hypersurface and the perturbation of its tension.
This can only be done by studying the high energy corrections
of a specific model. Nevertheless, we now want to provide an
argument why we think that a scale invariant spectrum is obtained in models
where the collapsing phase is characterized by $a \propto
(-\eta)^q$ with $q \ll 1$.

As we have seen in the above examples, the matching conditions are
fixed by $T$, given as some combination of $\Psi$ and $\Psi'$, and
determine $\Psi_+$ in terms of $\Psi_-$, $\Psi_-'$, and of the
surface stress perturbation $\de P_s$. The general result we are
about to derive is based on one important assumption, the
smallness of $T$, as given in Eq.~(\ref{assumption1}).  As explained
there, this assumption precisely limits the 'infrared power' of the
'new physics' needed to convert contraction into expansion.
As we have seen
[Eqs.~(\ref{assumption2}) and (\ref{assumption3})], as a consequence 
the extrinsic curvature and tension perturbations, $\tilde \delta K$
and $\tilde\delta P_s$, have the same $k$-dependence as $\Psi$ and $\Psi'$.

This  assumption fixes completely the final spectrum,
avoiding any arbitrariness such as the one found
in~\cite{patrickmartin} for the ekpyrotic scenario. Then, in
Eqs.~(\ref{m1}) and (\ref{m3}) the $k$-dependence is given entirely
in terms of  the coefficients $A$ and $B$. As a result, the
$k$-dependence of the coefficients $A_+$ and $B_+$ is a mixture
of the $k$-dependence of $A_-$ and $B_-$ given by Eqs.~(\ref{A-})
and (\ref{B-}),
\bea
A_+(k) &=& \al_A k^{-(1+\mu)} + \beta_A k^{-1+\mu}, \\
B_+(k) &=& \al_B k^{-(1+\mu)} + \beta_B k^{-1+\mu},
\eea
where the $\al$-terms come from the $A_-$-mode and the $\beta$-terms
come from the $B_-$-mode. According to our assumption, the
coefficients $\al_\bullet$ and $\beta_\bullet$ generically contain a
constant and positive powers of $k\eta_1$. The $A_+$-mode is decaying and 
we may neglect it
soon after the matching. Generically we expect, according to the
amplitudes of the $A_-$ and $B_-$-modes, that $\al_A$ and $\al_B$
are much larger than $\beta_A$ and $\beta_B$. Comparing the $A_-$
and $B_-$-modes we expect
\be
{\cal O}\left(\al k^{-1-\mu} \right) \sim {\cal O}\left(
  \left({k\eta_1}\right)^{-2\mu}\beta k^{-1+\mu}\right),
\ee
hence, for super horizon modes, $k\eta_1 \ll 1$, we expect
$\al k^{-1-\mu}\gg \beta k^{-1+\mu}$, as long as $\mu=q+1/2$ is
positive. Therefore, one typically inherits the spectrum of the
$\al$-terms in the radiation era, leading to
\be
{\cal P}_{\Psi}
= |\Psi|^2k^3 = |\al_B|^2k^{1-2\mu} \quad \left( \propto k^{n-1} \right)
 .\label{specgen}
\ee
In this generic situation, we obtain a scale invariant spectrum
$1 \simeq n=2-2\mu= 1-2q$ if $q$ is close to zero, as in 
the ekpyrotic and modified pre-big bang case.

Only if the matching conditions are such that the $\al_B$-term is
suppressed by a factor  smaller than $ (k\eta_1)^{2\mu}$, the
$\beta_B$-term comes to dominate and the spectrum becomes
\be
 {\cal P}_{\Psi} =
|\Psi|^2k^3 = |\beta_B|^2k^{1+2\mu} \quad \left(\propto k^{n-1}\right)
 .\label{specspec}
\ee
Then, the spectral index $n=2+2\mu=3+2q$ results.

As an estimate, for scales of
order the present Hubble parameter, relevant for the
perturbations in the cosmic microwave background, $k = k/a_\pm
\sim H_0$, and for $1/\eta_1 = 1/(a_\pm \eta_1) \sim 10^{17}$ GeV,
we have $k\eta_1\sim 10^{-59}$! Hence we typically expect the
$\beta$-terms to be about $10^{59}$ times smaller than the $\al$-terms
on cosmologically relevant scales, $\al_\bullet
k^{-1-\mu} \sim 10^{59} \beta_\bullet k^{-1+\mu}$.

For the constant energy hypersurface we have obtained $\al_B\equiv 0$
and hence the generic inequality $\al k^{-1-\mu}\gg
\beta k^{-1+\mu}$ is violated. But if the
matching hypersurface deviates by more than about $\sim 10^{-59}$ from
the $\rho=$ constant hypersurface, we expect the $A_-$-term,
$\alpha k^{-1-\mu}$, to dominate in the Bardeen potential and to
determine the final spectrum.

For a scalar field without
potential, as in the original pre-big bang model, we have $q=1/2$ which in the
'generic case' leads to a spectral index $n=1-2q=0$ and only under very
special matching conditions, like matching on the constant energy hypersurface
with $\de P_s\equiv 0$, the spectral index $n=4$ is obtained.

In the case of ordinary inflation, $q\sim -1$, where $\mu=1/2+q$ is
negative, the situation is quite different. There, the $A_-$-mode is
decaying and the Bardeen potential at the end of inflation is
dominated by the constant $B_-$-mode. Hence, we generically expect to
inherit in the radiation phase the spectral index from the $B_-$-mode with
$n=3+2q$, leading to a scale invariant spectrum for ordinary inflation,
$q\sim -1$. This is also the spectrum obtained when matching on the constant
energy  hypersurface.

In Ref.~\cite{pp}, a radiation dominated contracting phase is
connected smoothly to a radiation dominated expanding phase, via a
scalar field with negative energy density which comes to dominate
in the high curvature regime. Here a $n=-1$ spectrum of
perturbation is found with analytical arguments and via numerical
simulation. This agrees with our result. In this case, in fact,
$q=1$ and according to our argument 
we would generically expect $n=1-2q=-1$, as obtained in
Ref.~\cite{pp}. It is interesting to note that the matching
conditions of Ref.~\cite{pp} corresponds to the matching on the
hypersurfaces determined by $T=-\HH^{-1}\Psi$ from longitudinal
gauge. According to Eq.~(\ref{shorti}), this corresponds to the
gauge with $\tilde\de g_{ij}=0$, $i \neq j$, the `off-diagonal gauge', which
has also been considered in Ref.~\cite{Bru} as the gauge in which
perturbations remain small during the pre-big bang phase.

This is our main result: When matching a collapsing universe to an
expanding one, we expect the Bardeen potential in the expanding phase to
inherit the spectrum
of the mode which grows during the collapse phase, leading to
\be
{\cal P}_{\Psi} \propto k^{-2q} ,~~~ n=1-2q \label{genspec},
\ee
where $q$ is the exponent with which the scale factor contracts in
conformal time, $a\propto |\eta|^q$. Remind that this result holds
only if we assume, as explained in Sec.~\ref{matchsec}, that $T$
is small on large scales [see Eq.~(\ref{assumption1})].

\section{Using $\Psi$ or $\zeta$ ?}

In the above discussion we have used mainly the
Bardeen potential $\Psi$. Several authors~\cite{robi,hwang,hwang2,robi2}
use the curvature perturbation
$\zeta$ given in Eq.~(\ref{me1}).
In particular, Ref.~\cite{robi2} has found
\be
\zeta \propto {|\eta|^{1/2}\over a}H^{(2)}_\nu(k\eta), \qquad \nu = |q-1/2|.
\ee
This also follows from the definition of $\zeta$ [see
Eq.~(\ref{me1})], together with the solution (\ref{solPsi}) for
$\Psi$. During the pre-big bang phase, $\eta<-\eta_1$, this leads
to the following
 spectrum for $\zeta$ on super horizon scales, modulo logarithmic corrections,
\be
 {\cal P}_\zeta = |\zeta|^2k^3 \propto \left\{ \begin{array}{ll}
   k^{4-2q}|\eta|^{2-4q}  & \mbox{for }~ q > 1/2, \\
   k^{2+2q} & \mbox{for }~ q< 1/2 ,\end{array} \right.
\label{zetaspec}
\ee
giving a spectral index for the variable $\zeta$,
\be
  n_\zeta =  \left\{ \begin{array}{ll}
    5-2q   & \mbox{for }~ q > 1/2, \\
        3+2q   & \mbox{for }~ q< 1/2 .\end{array} \right.
\ee
Since in Ref.~\cite{robi2} the matching condition $[\zeta]_\pm=0$ is used,
the spectral index of $\zeta$    translates directly into the spectral index
of scalar perturbations in the radiation era, where $\zeta$ and $\Psi$
essentially agree on super horizon scales. This is the reason why these
authors obtain a scale invariant spectrum also for $q=2$ (while they
obtain $n=3$ for the ekpyrotic model).

We have found the following behavior of the $\Psi$ spectrum on
super horizon scales during the pre-big bang phase (see
Eq.~(\ref{solPsi}) in the limit $k|\eta| \ll 1$),
\be
 {\cal P}_\Psi  \propto \left\{ \begin{array}{ll}
   k^{-2q} |\eta|^{-(2+4q)}  & \mbox{for }~ q>-1/2 , \\
   k^{2+2q}   & \mbox{for }~ q < -1/2   .\end{array} \right.
\label{Psispec}
\ee
This leads to the spectral index of $\Psi$,
\be
  n_\Psi =  \left\{ \begin{array}{ll}
    1-2q   & \mbox{for }~ q> -1/2 , \\
        3+2q   & \mbox{for }~ q< -1/2 .\end{array} \right.
\ee
Comparing Eq.~(\ref{zetaspec}) and Eq.~(\ref{Psispec}) we see that
\be
{\cal P}_\zeta \simeq |k\eta|^{2\ga} {\cal P}_\Psi  \le {\cal P}_\Psi,
\ee
with
\be
  \gamma =  \left\{ \begin{array}{ll}
    0  & \mbox{for }~ q< -1/2, \\
    1+2q  & \mbox{for }~ -1/2<q<1/2,  \\
        2   & \mbox{for }~ q> 1/2  .\end{array} \right.
\ee

As we have mentioned above, for cosmologically relevant scales, the factor
$|k\eta|$ becomes of the order of $10^{-59}$ at the matching surface. We have
argued in the previous subsection that the larger variable $\Psi$ should
be relevant at the matching surface, and only under very special special
matching conditions the spectral index  of $\zeta$ is inherited after
the big bang. Generically we therefore expect $n=n_\Psi$ to be the
spectral index in the radiation era.  If $q<-1/2$, $\Psi$ and $\zeta$ 
agree up to a constant pre-factor, and this distinction becomes
irrelevant for the spectral index. This is exactly what happens in
'ordinary inflation' where $q\sim -1$. The functions $n_\Psi$ and
$n_\zeta$ are shown in  Fig.~1.

\begin{figure}[ht]
\centerline{\epsfig{file=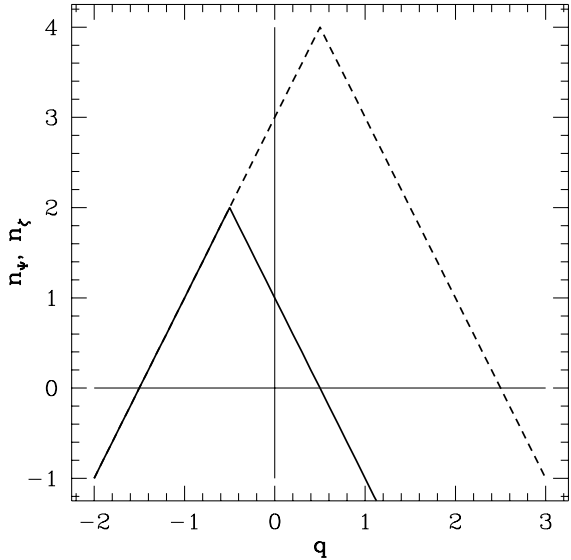, width=3.1in}}
\caption{The spectral indices $n_\Psi$ (solid) and $n_\zeta$ (dashed) are
shown as a function of $q =\HH \eta$. As argued in the text, we expect
a resulting spectral index $n=n_\Psi$ in the radiation era.}
\end{figure}

Finally, for completeness, we want to  emphasize that the Bardeen
potential in a radiation dominated universe really determines the
spectral index $n$ via $ {\cal P}_{\Psi} = |\Psi|^2k^3 \propto
k^{n-1}$. A scale invariant spectrum is defined as one for which
$\left\langle\left(\de M/M\right)^2\right\rangle_{\rm h.c.}$ is scale
independent, where the brackets denote spatial average and the
subscript ${\rm h.c.}$ indicates the scale of horizon crossing. Therefore,
the spectral index is defined by  $ \left\langle\left(\de
M/M\right)^2 \right\rangle_{\rm h.c.} \propto k^{n-1}$, so that $n=1$
represents a scale-invariant spectrum. 

On the other hand
\be
\left< \left( {\de M\over M}\right)^2 \right>
=k^3\left|{\de\rho\over \rho}\right|^2 .
\ee
On sub horizon scales and also at horizon crossing, $\de\rho/
\rho$ is not strongly gauge dependent, so we may choose whatever
gauge we please. We use comoving gauge (it is a simple estimate to
verify the same behavior, {\em e.g.}\ for longitudinal gauge). In
comoving gauge we have the constraint equation~\cite{ruth},
\be
k^2\Psi = {3\HH^2 \over 2}\left({\de\rho\over \rho}\right)_{\rm
com.} .
\ee
Using that $\HH\simeq k$ at horizon crossing and that
$\Psi$ is time independent on super horizon scales, we get
\be
  \left|\left({\de\rho\over \rho}\right)_{\rm com.}\right|^2_{\rm h.c.}
 \simeq |\Psi|^2 ,
\ee
hence
\[
\left\langle \left|\de M/M\right|^2 \right\rangle_{\rm h.c.} \simeq
    k^3|\Psi|^2 = {\cal P}_{\Psi} \quad \propto k^{n-1}.
\]
In the radiation dominated era $\zeta$ is roughly equal to $\Psi$ and the
above equation therefore holds also for ${\cal P}_\zeta$.

\section{Conclusions}
We have discussed the matching from a collapsing to an expanding
Friedmann universe. 
We have noted that a non-vanishing surface tension at the
matching surface is needed to turn the pre-big bang collapse into
expansion. This surface tension and its perturbation have to be
specified by the high energy corrections of the theory. It is this
surface tension which determines the correct matching surface and
it will generically not be parallel to the $\rho + \delta \rho=$
constant surfaces.

We have found that, if the matching is
performed at the $\rho + \delta \rho=$ constant hypersurface, the
growing mode from the pre-big bang phase is converted entirely
into the decaying mode in the radiation phase. In this case the
spectral index $n=3+2q$ is obtained, leading to $n=3$ for the
ekpyrotic and modified pre-big bang model, and $n=4$ for the original
pre-big bang model. However, if the matching hypersurface is
chosen to be somewhat different from $\rho + \delta \rho=$
constant, one  obtains $n=1-2q$. Hence, the ekpyrotic
and the modified pre-big bang model can lead to a scale
invariant spectrum of scalar perturbations. 

Our result is based on the assumption that perturbing
our background bouncing universe does not change completely the
time and duration of the bounce on large scales. We have formulated
this requirement precisely by restricting the allowed 'infrared power'
of $T$.

Notice that the
spectral index resulting from our matching conditions of a
pre-big bang transition, is never blue, $n \le 1$. This is not so
surprising: On sub-horizon scales, the perturbations are in their vacuum
state. They start growing as soon as they exit the horizon until
the end of the pre-big bang phase. Hence large scales, which exit
earlier, have more time to grow.

Often, as a heuristic approach to obtain the spectrum of fluctuations,
one considered $| \Psi|^2 k^3$ at horizon crossing requiring that
this behaves like $k^{n-1}$. Applying this procedure during the
pre-big bang at the first horizon crossing (exit), one obtains the
blue spectra $n=3$ for the ekpyrotic or the modified pre-big bang
model and $n=4$ for the original pre-big bang respectively.
However, if one determines the same quantity at the {\em second}
horizon crossing (re-entry), during the radiation dominated
phase, one obtains the correct spectral indices $n=1$ and $n=0$
respectively. Since in an expanding universe the Bardeen potential
does not grow on super horizon scales, it does not matter at which
horizon crossing, exit or re-entry, the spectrum is determined in
the case of ordinary inflation. In a pre-big bang model however, this
difference is crucial as we have seen.

The discussion presented in this paper does not affect the gravity wave
spectrum~\cite{gwaves} which still leads to
the spectral index $n_T=3$ for both models and is a potentially important
observable to discriminate them from ordinary inflation.

The main open problem when studying this bouncing models 
remains the high energy
transition from the pre- to the post-big bang. There, corrections
should become important, and we have assumed here that for super
horizon scales they can be summarized into a tension on the
matching surface. Furthermore, it has not yet been shown from
string theory that the dilaton can obtain an exponential potential
(in the modified pre-big bang model) or that the brane distance
simply obeys the equation of motion of a minimally coupled scalar
field with exponential potential from the brane point of view for
the ekpyrotic model.

Also the quantum production of other modes possible in these models,
{\em e.g.}~the
axions and moduli in the modified pre-big bang, or the 'graviphoton' and
'graviscalar' coming from the extra-dimension in the ekpyrotic model,
have to be investigated. 

Nevertheless, we conclude that models where high energy
corrections lead a slowly collapsing universe over into an
expanding radiation dominated phase may represent viable
alternatives to usual 'potential inflation', in generating a
scale invariant spectrum of perturbations. However, many open
questions, especially concerning the high energy corrections, and flatness,
still have to be properly addressed. \vspace{0.5cm}

{\bf Acknowledgment}~~~ We thank Robert Brandenberger, Cyril Cartier,
Fabio Finelli, Maurizio Gasperini, David Langlois, Andrei Linde,
Jer\^ome Martin, Patrick Peter, Jean-Philippe Uzan, and
Gabriele Veneziano for stimulating and clarifying discussions.
   
}
\end{document}